\documentclass[aps,prb,twocolumn,
				amsmath,amssymb,amsfonts,floatfix,
				longbibliography,superscriptaddress,nobibnotes]{revtex4-2}

\usepackage{graphicx} 
\usepackage{bm} 
\usepackage{ulem}
\usepackage{physics}
\usepackage{mathtools}
\usepackage[hypertexnames=false,colorlinks=true,citecolor=blue,
			linkcolor=blue,urlcolor=blue]{hyperref}

\usepackage{xfrac}

\usepackage{xcolor}

\DeclareUnicodeCharacter{2212}{\ensuremath{-}} 

\graphicspath{{fig/}{./fig/}{.}}

\begin{document}
\title{Topological in-gap chiral edge states \\in superconducting Haldane model with spin--orbit coupling}

\author{Sajid Sekh}
\email[e-mail: ]{sajid.sekh@ifj.edu.pl}
\affiliation{\mbox{Institute of Nuclear Physics, Polish Academy of Sciences, W. E. Radzikowskiego 152, PL-31342 Krak\'{o}w, Poland}}

\author{Andrzej Ptok}
\email[e-mail: ]{aptok@mmj.pl}
\affiliation{\mbox{Institute of Nuclear Physics, Polish Academy of Sciences, W. E. Radzikowskiego 152, PL-31342 Krak\'{o}w, Poland}}

\date{\today}

\begin{abstract}
Topological superconductivity is currently one of the prime interests, given the properties of its exotic nature of chiral edge states. A broken time-reversal symmetry (TRS) is an essential ingredient in the recipe of a chiral edge state. The Haldane model is one of the many factors that can break TRS in a system. Thus, we explore the possibility of topological superconductivity in the Haldane model under the influence of a conventional superconductor. The edge states originating from such recipes mostly remain outside the superconducting gap. Contrary to this, in the presence of spin-orbit coupling, the edge modes lie within the superconducting gap, and can lead to a gapless state for some range of parameters. Moreover, we use band inversion and projection on the real-space lattice to confirm the topological and chiral nature of the obtained edge states.
\end{abstract}

\maketitle

\section{Introduction}
\label{sec:intro}

Topological superconductivity (TSC) combines the idea of superconductivity and topological order. Given the electron-hole symmetry, the superconductors are a natural platform for exotic bound states like Majorana edge modes. These bound states are an example of a chiral surface/edge state with unidirectional propagation, which is observed in superconductors with broken time-reversal symmetry (TRS). Experimentally, a number of techniques like muon spin rotation $\mu$SR, neutron scattering, and polar Kerr effect can probe the lack of TRS.

The absence of TRS is a crucial step in the formation of exotic chiral states, and applying a magnetic field has been a popular choice for breaking the symmetry. This is because theoretical blueprints~\cite{takahashi.sato.09,lutchyn.sau.10} predict that chiral edge states appear if a magnetic field is applied to a system with {\it s}-wave pairing and Rashba spin--orbit coupling (SOC). However, a broken TRS state can also arise from other factors like unconventional pairing symmetry~\cite{avers.gannon.20,pustogow.luo.19,matsura.masakai.23,singh.scheurer.20,sigrist.ueda.91}, charge density order~\cite{mook.sidis.08,yu.wang.21}, orbital current loops~\cite{fauque.sidis.05,li.baledent.10,verma.97}, intrinsic magnetic order~\cite{luke.fudamoto.98,kaminski.rosenkranz.02}, or due to a combination of aforementioned factors~\cite{mielke.das.22,deng.liu.24,shang.zhao.23,mandal.kataria.24}. In addition, twisted bilayer cuprate~\cite{can.tummuru.21} is predicted to feature chiral edge modes, where twisting of the layers gives rise to $d\pm id^{\prime}$ order parameter that breaks TRS.

The realization of chiral topological states, and their potential application in fault-tolerant quantum computing~\cite{sarma.freedman.15,vijay.hsieh.15} makes studying TSCs highly interesting in present times. So far, a number of diverse platforms have been proposed to realize TSC, which include semiconductor--superconductor heterojunction~\cite{lutchyn.sau.10}, helical Shiba chains~\cite{schneider.beck.21}, iron-based superconductors~\cite{wang.kong.18}, transition metal dichalcogenides~\cite{hsu.vaezi.17}, and skyrmionic magnons~\cite{maeland.subdo.23} smoking gun evidence for chiral edge states remains highly challenging given factors like the sophistication of experiments, fine-tuning of parameters, and material quality.

\begin{figure}[t]
	\includegraphics[width=\columnwidth]{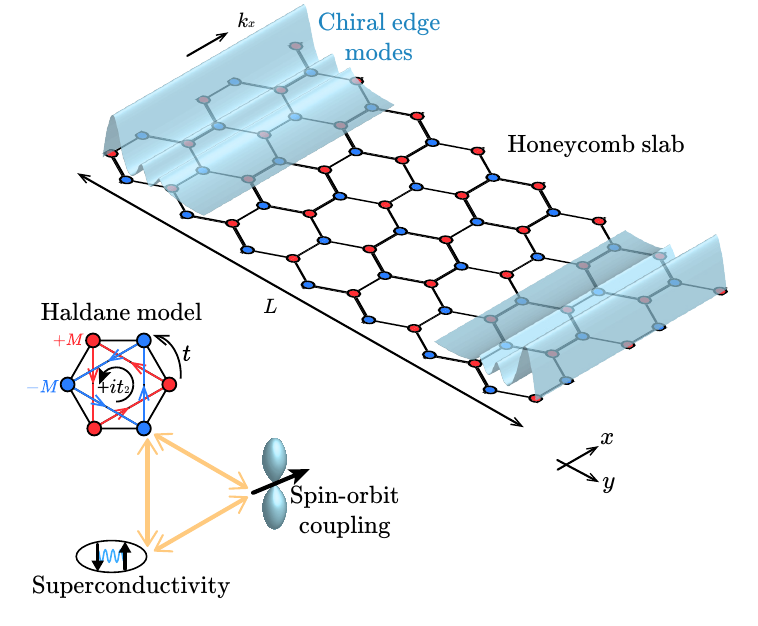}
	\caption{The schematic shows a zigzag honeycomb slab of length $L$ with periodic boundary condition along $x$ and open boundary condition along $y$ direction. We consider the Haldane model in the presence of superconductivity and Rashba spin--orbit coupling. A combination of these leads to subgap chiral edge states, which are localized on the boundary of the slab.}
	\label{fig:schematic}
\end{figure}

\begin{figure*}[t]
	\includegraphics[width=\linewidth]{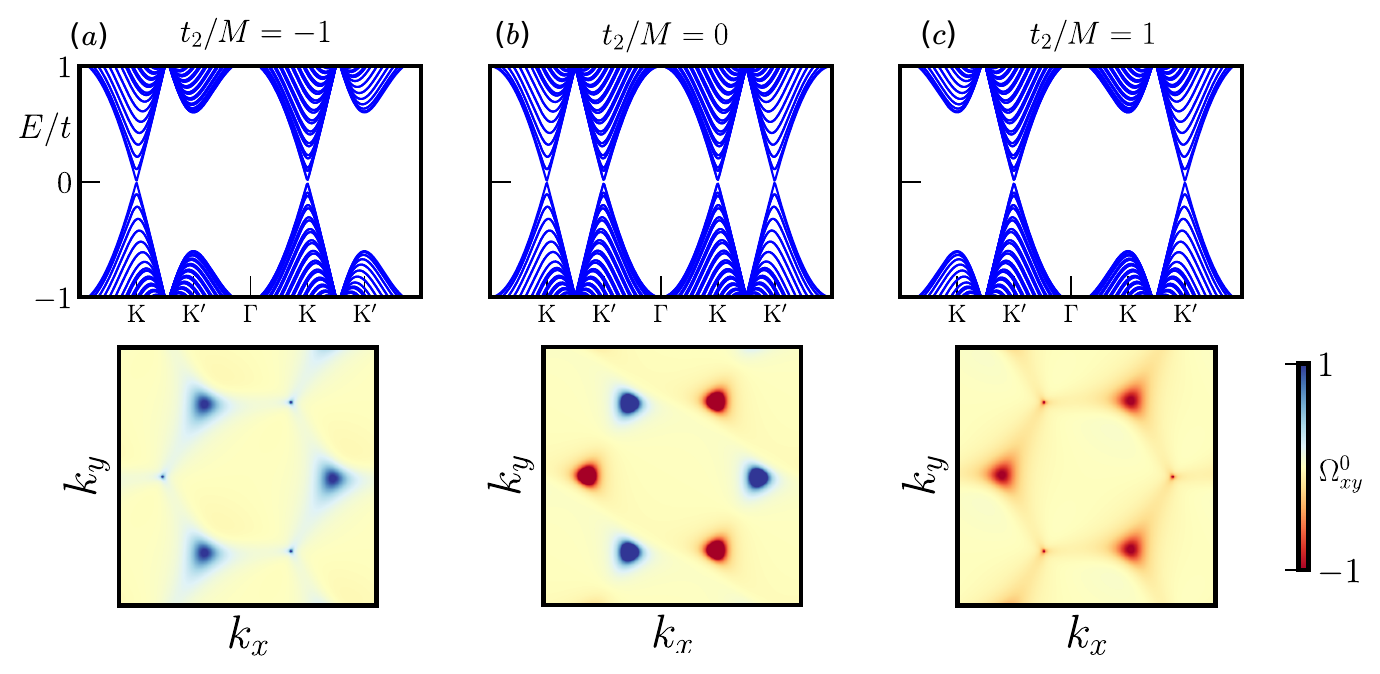}
	\caption{The figure highlights the role of the mass term $M$ and next-nearest neighbor complex hopping $t_{2}$ in the topological features of the Haldane model. 
The band structure (top) and the corresponding Berry curvature in the reciprocal space (bottom) are shown for the different Haldane parameters $t_{2}/M$ (as labeled).}
	\label{fig:analytical}
\end{figure*}

The fate of a superconducting state strongly depends on the inversion and TRS. The absence of inversion in noncentrosymmetric superconductors~\cite{frigeri.agterberg.04,dimitrova.feigelman.07,fischer.sigrist.23} facilitates SOC. This leads to unusual features like enhanced Pauli limit~\cite{gorkov.rashba.01} and singlet-triplet mixing~\cite{frigeri.agterberg.04,edelshtein.89}. Additionally, if the TRS is broken by a magnetic field, finite momentum pairing state~\cite{barzykin.gorkov.02} can arise, which are stabilized by Lifshitz invaraints~\cite{agterberg.11} in the free energy.

In honeycomb lattices, it has been shown that Rashba SOC and magnetic field~\cite{min.hill.06,qiao.yang.10} produces a gapped spectrum with finite Berry curvature. When {\it s}-wave superconductivity is added, the system features zero-energy Majorana edge states at the interface~\cite{rainis.trifunovic.13,beenakker.13,jose.lado.15,sato.takahashi.09} or inside the vortex~\cite{qi.hughes.10}.

It is evident that a broken TRS is crucial for obtaining chiral topological phases. Apart from the magnetic field and unconventional pairing, the Haldane model~\cite{haldane.88} can also break the TRS. It describes a Chern insulator that exhibits the quantum anomalous Hall effect — a phenomenon analogous to the quantum Hall effect~\cite{klitzing.dorda.80} but without the need for a magnetic field. While similar to the Kane--Mele model~\cite{fu.kane.08}, the Haldane model applies specifically to spinless fermions. In essence, it consists of three terms: nearest neighbor (NN) hopping, the next-nearest neighbor (NNN) complex hopping, and a staggered flux piercing the honeycomb plaquette. Here, TRS is broken due to the nature of NNN hopping since it is either positive or negative depending on the direction. Originally, it was shown~\cite{haldane.88} that a trivial system can be rendered topological by tuning the ratio of the complex hopping and staggered flux. The topological phase of the Haldane model is characterized by the Chern number $C=\pm 1$. Thus, the question naturally follows: Is it possible to realize a TSC by pairing the topological edge states of the Haldane model with conventional superconductivity? Thus, we consider the Haldane model in presence of the superconductivity and Rashba SOC (Fig.~\ref{fig:schematic}). Our detailed investigation reveals that, with such a recipe, the edge states always remain outside the uniform {\it s}-wave superconducting gap. However, we find that a Rashba SOC can produce tunable in-gap edge states that are chiral in nature.

We organise the paper as follows. First, we write down the tight-binding Haldane model in momentum space to discuss its energy landscape and topological properties in Sec.~\ref{sec:haldane}. Since momentum space description is incapable of producing edge states, we visualize the topological edge states in real space using a local marker in Sec.~\ref{ssec:chern_marker}. We discuss the superconducting Haldane model and the effect of SOC in Sec.~\ref{sec:haldane_sc}. In particular, we describe the Bogoliubov-de Gennes formalism for superconductivity in the presence of SOC in Sec.~\ref{ssec:bdg}. Then we present our comprehensive results discussing the effects of superconductivity and SOC (Sec.~\ref{ssec:sc} and Sec.~\ref{ssec:soc}, respectively). At last, we conclude the paper in Sec.~\ref{sec:summary}.

\section{Properties of the pristine Haldane model}
\label{sec:haldane}

\subsection{Tight-binding description}
\label{ssec:haldane_pristine}

The Haldane model consists of the NN hopping of electrons on a honeycomb lattice, followed by a NNN complex hopping. This can be captured by a tight-binding Hamiltonian as shown below
\begin{eqnarray} \label{eq:ham_normal}
	H_0 = H_T + H_h.
\end{eqnarray}
Here, $H_T$ is given by
\begin{eqnarray}
	H_T = -t\sum_{\langle ij \rangle} c^{\dagger}_{i\sigma} c_{j\sigma} - \mu \sum_i c^{\dagger}_{i\sigma} c_{i\sigma},
\end{eqnarray}
with $t$ and $\mu$ referring to the NN hopping integral and chemical potential, respectively. The fermionic operators $c^{\dagger}_{i\sigma} (c_{i\sigma})$ denote creation (annihilation) of electrons at the $i$-th site with spin $\sigma$: ${\bm r}_i = {\bm R}_i + {\bm \tau}$. Here, ${\bm R}_i = m {\bm a}_1 + n {\bm a}_2$ (for $ m,n\in\mathbb{Z}$) denotes the positions of the atoms in the unit cell, with ${\bm a}_{1,2}$ as a lattice vectors, and ${\bm \tau}$ represents the position of A or B sublattice atom within the unit cell.

\begin{figure}[t]
	\includegraphics[width=\linewidth]{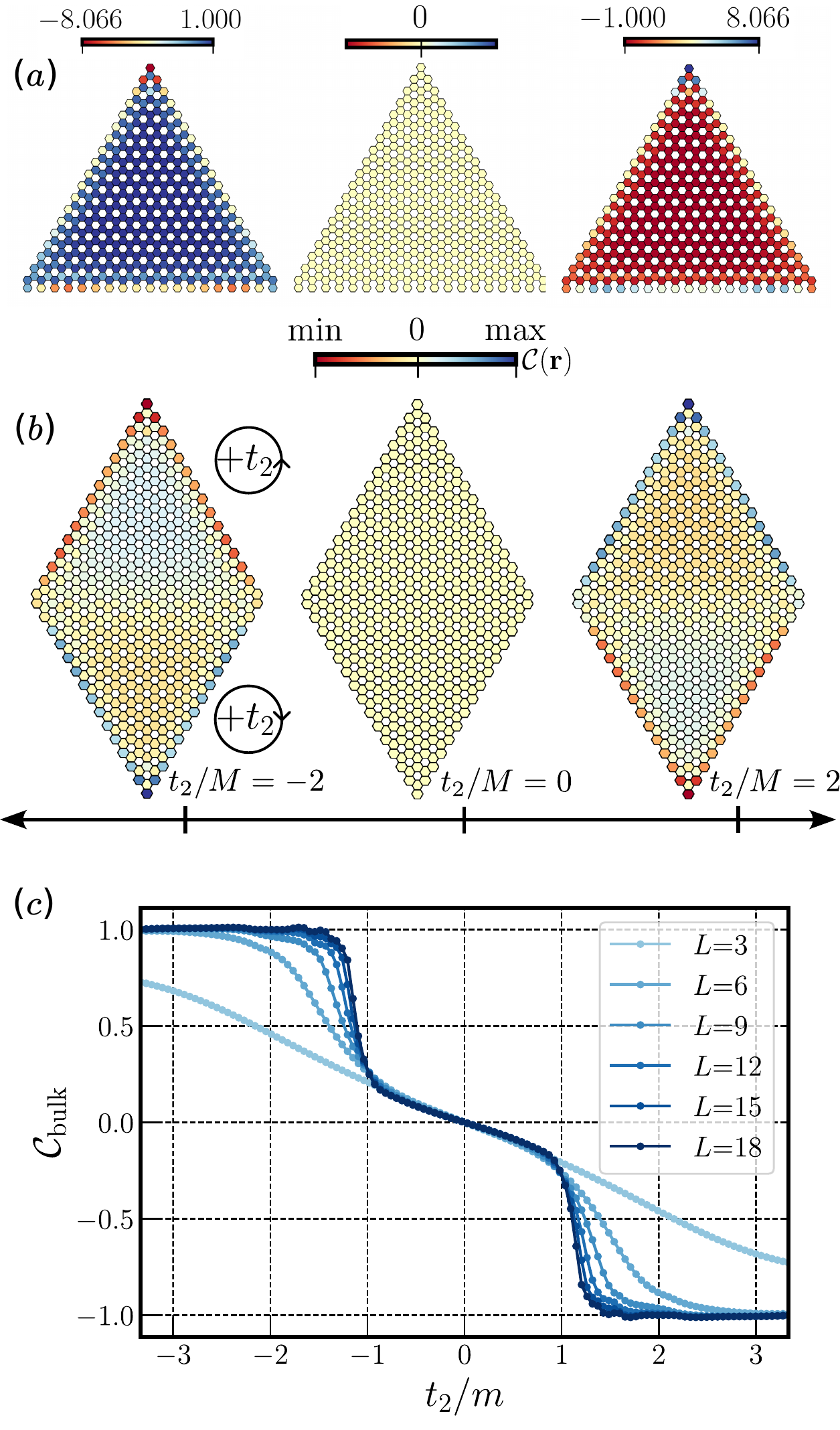}
	\caption{(a) The local Chern marker $\mathcal{C}(\mathbf{r})$ is projected on (a) triangular and (b) diamond honeycomb islands for $t_2/M=-2, 0, 2$. The diamond island is formed by combining two triangles presented on top panel, where the upper and lower parts have opposite chirality $t_{2}$ (c) The bulk local Chern marker $\mathcal{C}_{\text{bulk}}$ is plotted against $t_2/M$ for different system size $L$. Here, the Haldane parameter $t_2/M$ is taken as labeled. In all of the panels, we set $\mu/t=0$.}
\label{fig:chern_marker}
\end{figure}

The next term $H_h$ refers to the Haldane Hamiltonian, and reads as
\begin{eqnarray}
	H_h= -\frac{t_2}{3\sqrt{3}} \sum_{\langle\langle ij \rangle\rangle} e^{\phi_{ij}} c^{\dagger}_{i\sigma} c_{i\sigma} + \sum_{i\sigma} (-M)^i c^{\dagger}_{i\sigma} c_{i\sigma} ,
\end{eqnarray} 
where $t_2$ is the NNN hopping integral. The flux $\phi_{ij}$ piercing the honeycomb plaquette has a chiral character depending on the direction of hopping. That is, it is set to either $+\phi$ if the hopping is clockwise or $-\phi$ if the hopping is anticlockwise, with respect to the center of the plaquette. Following the proposal of Haldane~\cite{haldane.88}, we take $\phi=\pi/2$ throughout our calculation so that NNN hopping is purely complex. Furthermore, $M$ refers to a staggered on-site potential, which takes the value of $+M$ or $-M$ depending on the sublattice (see Fig.~\ref{fig:schematic}).

\subsection{Topological features}
\label{ssec:berry_curv}

In the reciprocal space, the Hamiltonian can be written in a $2 \times 2$ matrix form:
\begin{eqnarray}
	H_0(\mathbf{k}) = h_x \tau_x + h_y \tau_y + h_z \tau_z,
\end{eqnarray}
where
\begin{eqnarray}
	\nonumber h_x(\mathbf{k}) &=& t \left[ 1+\cos \left( \mathbf{k}\cdot \mathbf{a}_1 \right) + \cos \left( \mathbf{k}\cdot \mathbf{a}_2\right) \right] , \\
	h_y(\mathbf{k}) &=& t \left[ \sin \left( \mathbf{k}\cdot\mathbf{a}_1 \right) + \sin \left( \mathbf{k}\cdot\mathbf{a}_2 \right) \right] , \\
	\nonumber h_z(\mathbf{k}) &=& M + 2t_2 \sin \phi \left[ \sin \left( \mathbf{k} \cdot \mathbf{a}_1 \right) - \sin \left( \mathbf{k} \cdot \mathbf{a}_2 \right) \right. \\ 
	\nonumber && \left. - \sin \left( \mathbf{k} \cdot \mathbf{a}_1-\mathbf{k} \cdot \mathbf{a}_2 \right) \right] .
\end{eqnarray}
Here $\tau_i$ ($i=x,y,z$) refers to the Pauli matrices acting on the sublattice space, and $h_i$ are its coefficients. Note that, the model, in essence, is tuned by two parameters: $M$ and $t_2$. The system is in a topologically trivial state for $-1<t_2/M<1$. Tuning the ratio outside this window can drive the system into a topological state with Chern number $+1$ or $-1$. The topological nature is easy to follow from the Berry curvature
\begin{eqnarray}
	\nonumber \Omega^{n}_{xy} = i \sum_{n'\neq n} \frac{\langle n \vert \partial_{k_x} H \vert n'\rangle \langle n' \vert \partial_{k_y} H \vert n\rangle - (k_x \xleftrightarrow{} k_y)}{(\epsilon_{n} - \epsilon_{n'})^2} . \\
\end{eqnarray}

We present the band structure and Berry curvature of the continuum model in Fig.~\ref{fig:analytical}. In momentum space, the pristine honeycomb lattice exhibits TRS and three-fold rotational symmetry $C_3$, among other symmetries. Both of these symmetries are present when $t_2/M=0$. This is because, at K and K' points, the bands touch each other, and the Berry curvature takes opposite signs since TRS requires $\Omega(\mathbf{k})=-\Omega(-\mathbf{k})$. These features are seen in Fig.~\ref{fig:analytical}(b). Contrary to this, setting $\vert t_2/M \vert=1$ breaks both symmetries. Particularly, a gap opens either at K or K' points depending on $\mathrm{sgn}(t_2/M)$, and the Berry curvature weight shifts from one valley to the other. As a result, the system has a net Berry curvature with Chern number $+1$ or $-1$.

\begin{figure*}[]
	\includegraphics[width=\linewidth]{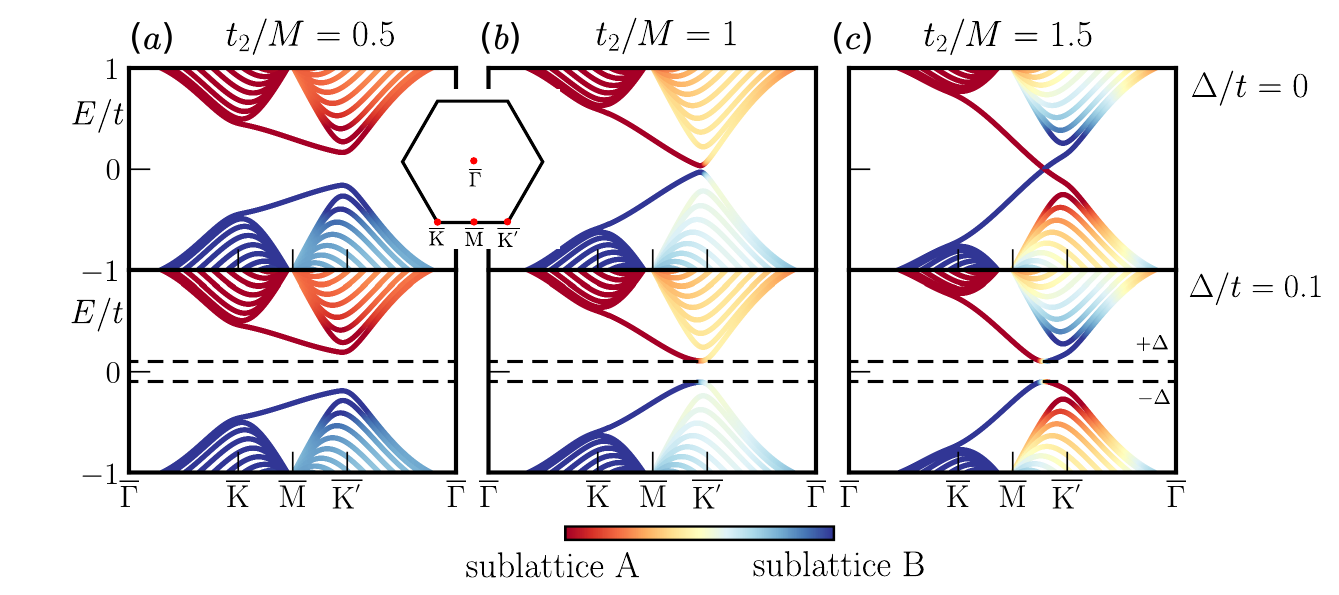}
	\caption{The sublattice projected band structure is shown along the high symmetry path of the Brillouin zone (presented in inset) for different Haldane parameters $t_{2}/M$ (as labeled).
Here, colors correspond to the contribution from sublattices A and B.
The top and bottom panels represent results for the Haldane model in the absence and presence of superconductivity, respectively.
The superconducting gap is shown by the dashed lines.
Results are obtained in the absence of SOC and $\mu / t = 0$.}
	\label{fig:slab}
\end{figure*}

\subsection{Real-space Chern marker}
\label{ssec:chern_marker}
The non-triviality of an edge state is characterized by a topological invariant, which is defined for an infinite lattice. In fact, the bulk properties and edge states are deeply intertwined according to the bulk-boundary correspondence~\cite {fukui.shiozaki.12,mong.shivamoggi.11}. To visualize this bulk-edge relationship, we compute the local Chern marker~\cite{bianco.resta.11} (LCM) of a finite honeycomb system. This maps the Chern number to real space by computing a local quantity, which is given by
\begin{eqnarray}
	\mathcal{C}(\mathbf{r}) = -\frac{4\pi}{A} \langle \mathbf{r} \vert \hat{P} [\hat{x},\hat{P}] [\hat{y},\hat{P}] \vert \mathbf{r}\rangle .
\end{eqnarray}
Here, $A$ is the area of the unit cell, $\vert \mathbf{r}\rangle$ is the position operator with $\hat{x}$ and $\hat{y}$ being its $x$ and $y$ components, respectively. Furthermore, $\hat{P}$ is the projection operator on the occupied states
\begin{eqnarray}
	\hat{P} = \sum_{n \in \{\epsilon_n<0\}} \vert \psi_{n}\rangle \langle \psi_{n} \vert .
\end{eqnarray}
The LCM can be written in terms of the bulk and edge contributions as $\mathcal{C}(\mathbf{r}) = \mathcal{C}_{\text{bulk}} + \mathcal{C}_{\text{edge}}$. The bulk part remains mostly uniform, and reflects the Chern number. However, the projector $\hat{P}$ changes at the boundary because of gapped edge states, causing the edge marker to fluctuate. Additionally, compared to bulk, the LCM near edge tends to be larger in magnitude with an opposite sign, since a global trace of the local marker vanishes. This can be seen in Fig.~\ref{fig:chern_marker}(a), where we project LCM on triangular honeycomb islands. In such finite systems, the LCM is non-zero and uniform throughout the bulk. Moreover, in the topological phase, the LCM yields $-1$ for $t_2/M=2$ and $1$ for $t_2/M=-2$, which reflects the Chern number of the Haldane model. For $t_2/M=0$, the LCM vanishes everywhere, suggesting a trivial phase. If the system is small enough, the LCM might be dominated by edge fluctuations and finite-size effects. The bulk LCM will deviate from the Chern number in such a scenario. We capture such effet in Fig.~\ref{fig:chern_marker}(c), where we plot the $\mathcal{C}_{\text{bulk}}$ versus $t_2/M$ for varying system sizes. The result shows that $\mathcal{C}_{\text{bulk}}$ is quantzied with the magnitude $\pm 1, 0$. The transition to different quantized levels occurs at $|t_2/M|=1$ in accordance with the Haldane model. Notably, the transitions become sharper as the system size increases. For small sizes, the edge has a significant contribution to $\mathcal{C}_{\text{bulk}}$ as explained earlier, which causes the bulk marker to deviate. This deviation is most prominent near the transition.

With this result, an interesting question naturally arises: what happens at the interface when two triangular islands are connected? Therefore, in the next panels, we construct a diamond island by combining two triangular islands considered earlier. The novelty is that the hopping chirality is taken opposite in the top and bottom parts of the geometry. In practice, one can render such an opposite sense of chirality by inverting the flux arrangement in the nano-plaquettes.

Therefore, we systematically set the Haldane parameter $t_2/M$ to either $0$ or $\pm 2$, and showcase our results in Fig.~\ref{fig:chern_marker}(b). First, the case $t_2/M=0$ indicates a trivial phase -- this is correctly captured by the marker as LCM vanishes throughout the system. When $t_2/M$ is finite, specifically with an absolute value larger than $1$, the system exhibits a small LCM in the bulk and a large and opposite LCM at the edge. Having the opposite chirality convention explained earlier, the sign of LCM flips as one moves from the upper to lower region of the nanoflake and vice versa. Particularly, say for $t_2/M=2$, the bulk LCM in the upper region is close to $-1$, which changes to 1 in the lower region. Therefore, the bulk LCM of each region mimics the global Chern number, and is representative of different topological phases of the Haldane model. At the interface of upper and lower regions, the LCM goes to zero, similar to the $t_2/M=0$ scenario. This is due to the cancellation of opposite NNN hoppings at the interface, and is apt because transitioning from a topological state with $\mathcal{C}=-1$ to $\mathcal{C}=1$ requires a trivial phase in the middle with $\mathcal{C}=0$. Experimentally, such nontrivial edge modes and their chirality can be probed using non-local conductance measurements~\cite {ikegaya.asano.19,ptok.alspaugh.20}.

\section{Superconducting Haldane model with spin--orbit coupling}
\label{sec:haldane_sc}

\subsection{Rashba coupling and Bogoliubov--de Gennes formalism}
\label{ssec:bdg}

The topological nature of edge states is strongly influenced by factors like SOC and superconductivity. We investigate such effects by considering a honeycomb slab with periodic boundary conditions along the x-axis and open boundary conditions along the y-axis. In this way, we can obtain the bands along the $k_x$ direction with the footprints of edge modes arising from the zigzag open boundary [see Fig.~\ref{fig:schematic}]. Thus we write the Hamiltonian 
\begin{eqnarray} \label{eq:ham_new}
	H=H_{0} + H_{\text{soc}} + H_{\text{sc}}. 
\end{eqnarray} 
Here, the first term refers to the pristine Haldane model, which is introduced in Eq.~(\ref{eq:ham_normal}). 
The second term describes Rashba SOC along the out-of-plane direction $\hat{e}_z$\cite{ptok.glodzik.17}
\begin{eqnarray}
H_{\text{soc}}= -i\lambda_R \sum_{\langle ij \rangle} c^{\dagger}_{i\sigma} \left[ \left( {\bm d} \times {\bm \sigma} \right)_{\sigma\sigma'} \cdot \hat{e}_z \right] c_{j\sigma^{\prime}},
\end{eqnarray}
where $\mathbf{d}=\mathbf{r}_i-\mathbf{r}_j$ denotes the vector connecting two NN sites, $\boldsymbol{\sigma}=(\sigma_x,\sigma_y,\sigma_z)$ is the vector containing Pauli matrices acting in spin space, and $\lambda_R$ is the strength of the Rashba coupling. Note that, the SOC breaks the inversion symmetry along the z direction, and is responsible for mixing the $\uparrow$ and $\downarrow$ spin subspaces.

\begin{figure*}
	\includegraphics[width=\linewidth]{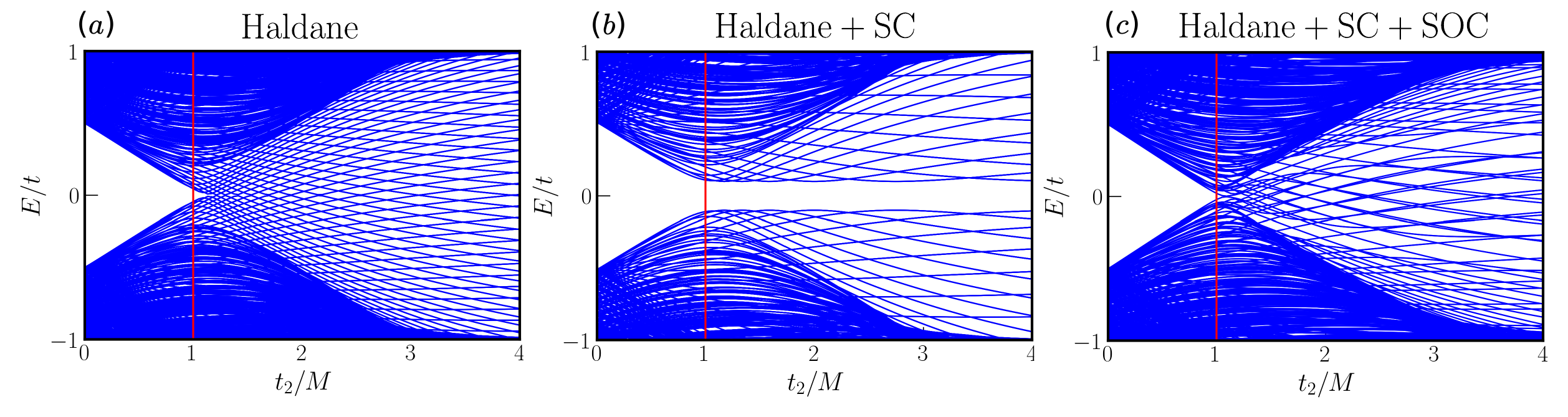}
	\caption{The energy spectrum of the slab is plotted against $t_2/M$ for three scenarios: (a) the ``simple'' Haldane model ($\Delta/t=0, \lambda_R/t=0$), (b) the Halane model with superconductivity ($\Delta/t=0.1, \lambda_R/t=0$), and (c) the Haldane model with superconductivity and Rashba SOC ($\Delta/t=0.1, \lambda_R/t=0.5$). Here, the red line corresponds to the topological phase transition point i.e. $t_2/M=1$ in the Haldane model. The results are shown for a slab of length $L=20$ unit. }
	\label{fig:energy_levels}
\end{figure*}

The last term introduces uniform {\it s}-wave superconductivity as follows
\begin{eqnarray}
	H_{\mathrm{sc}} = \Delta \sum_i (c^{\dagger}_{i\uparrow}c^{\dagger}_{i\downarrow} + c_{i\uparrow}c_{i\downarrow} ),
\end{eqnarray}
where $\Delta$ is the superconducting order parameter. The pairing Hamiltonian is not diagonal in this basis, which necessitates the introduction of the Bogoliubov transformation below
\begin{eqnarray}
	c_{i\sigma}=\sum_n (u_{in\sigma}\gamma_{n\sigma} - v^{\ast}_{in\sigma}\gamma^{\dagger}_{n\bar{\sigma}}).
\end{eqnarray}
In this description, the operators $\gamma_{n\sigma}$ ($\gamma^{\dagger}_{n\bar{\sigma}}$) annihilate (create) a Bogoliubov quasiparticle at $n$th state with spin $\sigma$, and $(u,v)$ denote the spectral weights. Such transformation leads to the Bogoliubov--de Gennes (BdG) equations~\cite{deGennes.97} in the form:
\begin{eqnarray}
\nonumber
E_{n}
\begin{pmatrix}
u_{in\uparrow} \\
v_{in\downarrow} \\
u_{in\downarrow} \\
v_{in\uparrow}
\end{pmatrix}
= \sum_j
\begin{pmatrix}
H_{ij\uparrow} & D_{ij} & S_{ij}^{\uparrow\downarrow} & 0 \\
D^{\ast}_{ij} & -H^{\ast}_{ij\downarrow} & 0 & S_{ij}^{\downarrow\uparrow} \\
S_{ij}^{\downarrow\uparrow} & 0 & H_{ij\downarrow} & D_{ij} \\
0 & S_{ij}^{\uparrow\downarrow} & D^{\ast}_{ij} & -H^{\ast}_{ij\downarrow}
\end{pmatrix}
\begin{pmatrix}
u_{jn\uparrow} \\
v_{jn\downarrow} \\
u_{jn\downarrow} \\
v_{jn\uparrow}
\end{pmatrix}. \\ 
\label{eq.bdg}
\end{eqnarray}
Here, $H_{ij\sigma}=-t\delta_{\langle ij \rangle \sigma}-t_2\delta_{\langle\langle ij \rangle\rangle \sigma}-\tilde{\mu}_{i\sigma}\delta_{ij}$ is the Haldane part, $S_{ij}^{\sigma\sigma^{\prime}}=-i\lambda_R \sum_{\langle ij \rangle} \big(\mathbf{d}_{ij}\times \boldsymbol{\sigma}^{\sigma\sigma^{\prime}} \big)_z $ corresponds to the Rashba coupling, and $D_{ij}=\Delta_i\delta_{ij}$ is the superconducting pairing term.

\begin{figure*}
	\includegraphics[width=\linewidth]{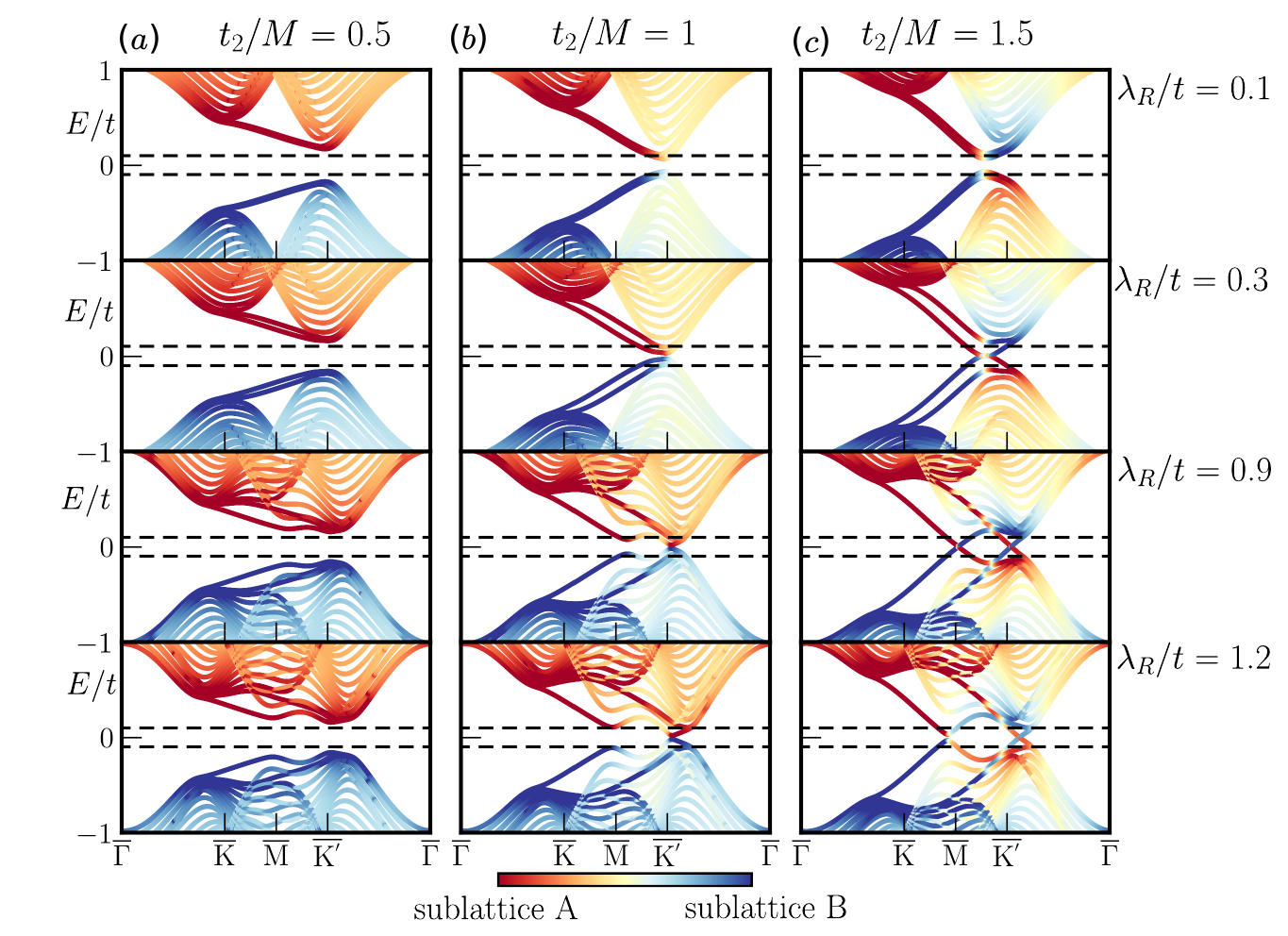}
	\caption{The sublattice projected band structure for the Haldane model in the presence of the superconductivity and Rashba SOC.
The colors correspond to the contribution of sublattice A and B. Results are obtained for different Haldane parameters $t_{2} / M$ (column-wise) and SOC $\lambda_{R}$ (row-wise) as labeled.
We take fixed superconducting order parameter $\Delta / t = 0.1$, and $\mu / t = 0$ in all panels.}
	\label{fig:soc}
\end{figure*}

\begin{figure}[t]
	\includegraphics[width=\linewidth]{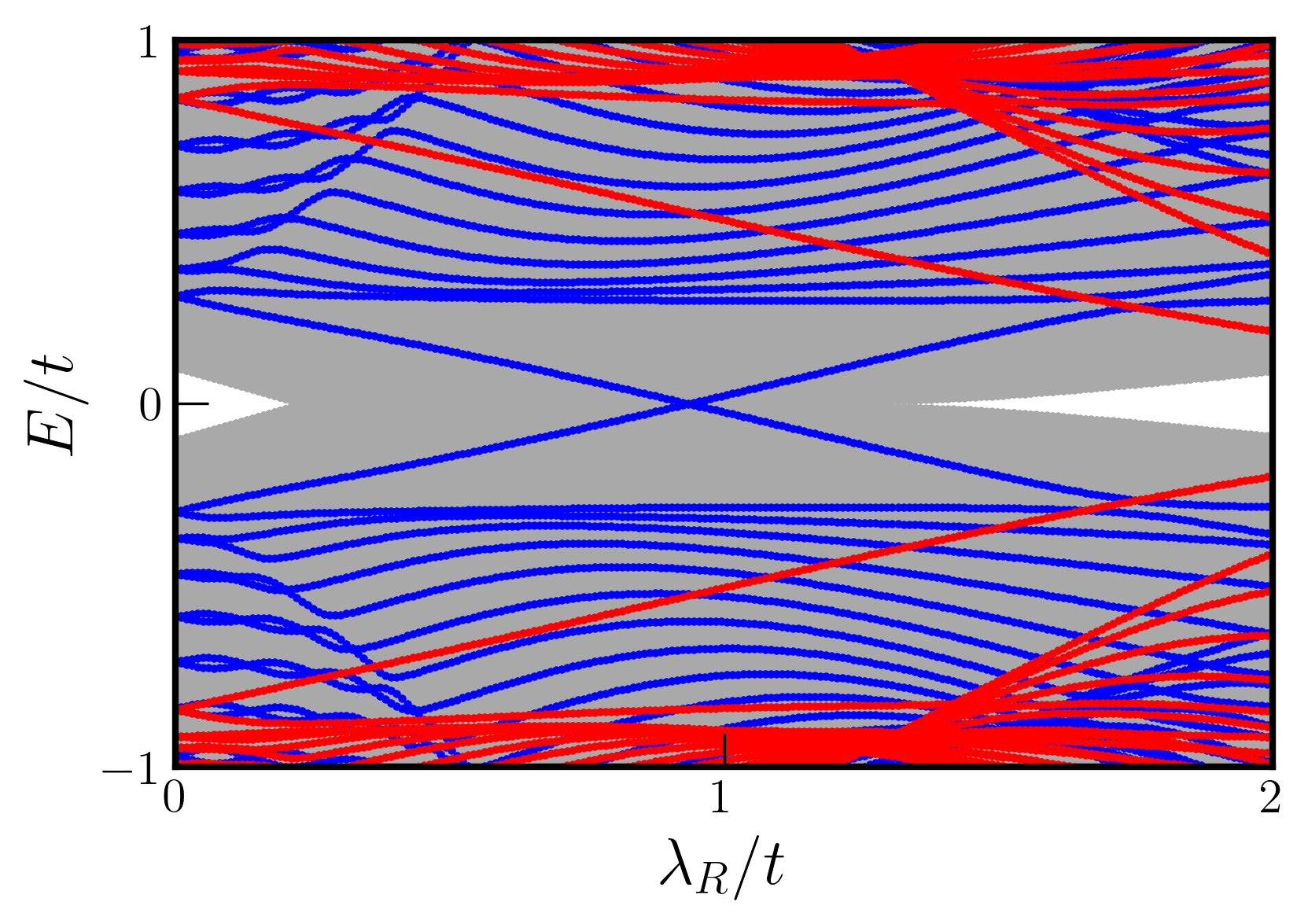}
	\caption{The energy spectrum is shown as a function of Rashba coupling $\lambda_R$ for the Haldane model with superconductivity and SOC.
	The red and blue colors correspond to the states at K and K' point, respectively, while other states are kept gray.
	The results are obtained for $t_2/M=2$, $\mu/t=0$, and $\Delta/t=0.1$.}
	\label{fig:evslambda}
\end{figure}

During our investigation, we computed several observables, which can be directly benchmarked with experimental techniques like scanning tunnelling microscopy (STM). We also calculate the local density of states (LDOS), which is useful in several instances
\begin{eqnarray}
	\nonumber \rho_{i\sigma}(\omega) = \sum_{n} \vert u_{i n\sigma}\vert ^2 \delta(\omega-E_{n}) + \vert v_{i n\sigma}\vert ^2 \delta(\omega+E_{n}), \\
\end{eqnarray}
where the summation runs over the state index $n$, $i$ is the site index, and $\sigma$ is the spin index. This is useful for projecting various degrees of freedom on the band structure, which can reveal important insights about the system.

Since we are considering the slab geometry, we have broken periodicity along $y$ direction. But the periodic boundary condition along $x$ direction is still intact, which makes $k_x$ a well-defined quantum number. This enables us to perform a Fourier transformation along one direction from real to momentum space: ${\bm r}_{i} = ( x_{i} , y_{i} ) \rightarrow ( k_{x} , y_{i} )$. Such transformation conserves position along $y$ direction, but transforms the $x$ coordinates. As a result, the BdG equations~\eqref{eq.bdg} are now a function of momentum -- which we exactly diagonalize to obtain the band structure along $k_x$ direction.

\subsection{Interplay between Haldane model and superconductivity}
\label{ssec:sc}

Let us now discuss how superconductivity affects features of the Haldane model. We present the band structure in the absence and presence of superconductivity in Fig.~\ref{fig:slab}. 

We start by discussing the Haldane model without superconductivity (see top panel of Fig.~\ref{fig:slab}). The pristine honeycomb band structure (not shown) features a pair of gapless Dirac cones at the K and K', which are connected by a zero-energy flat edge state. The system becomes gapped if $t_2/M$ is finite, and breaks TRS. Now the introduction of a small, non-zero Haldane parameter $\vert t_2 /M \vert < 1$ opens a gap in the spectrum [Fig.~\ref{fig:slab}(a)].

In the limit $t_2 \rightarrow M$, the gap starts to close at one of the valleys, and the system exhibits a topological phase transition at $t_2/M=1$. For $\vert t_2/M \vert > 1$, the system realizes gapless topological phases, and hosts a pair of edge modes with Chern number $\pm 1$. These states are localized at the boundary of the slab. The topological features of this transition can be evaluated by projecting the sublattice weights on the
bands. During the transition, the band character changes
significantly at K and K’ points. For instance, the sublattice character of the bands at the K’ point is flipped compared to its state before the topological phase transition [cf.~top panel of
Fig.~\ref{fig:slab}(a) and Fig.~\ref{fig:slab}(c)].

Similar results are found in the presence of superconductivity (bottom panels in Fig.~\ref{fig:slab}). However, contrary to the case without superconductivity, the topological phase transition does not close the energy gap. Although we do not find any in-gap states, we still observe the band-inversion in the topological phase, signifying the non-trivial character of the edge modes. In practice, all states are shifted from the Fermi ``zero'' level, as
expected from the BCS formula: $\pm \sqrt{\omega^2+\Delta^2}$, where $\omega$ denotes the energy of the states in absence of superconductivity. Due to the absence of gap closing in the superconducting state, the current scenario can be treated as an example of gapped topological edge modes in a trivial superconducting state.

\subsection{Effects of spin--orbit coupling}
\label{ssec:soc}

Previous works~\cite{takahashi.sato.09,lutchyn.sau.10,ptok.rodriguez.18} suggest that the topological character of a superconducting state can be altered by introducing the SOC. This is because SOC effectively mixes the spin subspace, which effectively gives rise to the Cooper pairs formed by spinless quasiparticles~\cite{das.ronen.12}, similarly to the Kitaev model~\cite{kitaev.01}. Here, the topological transition from trivial to non-trivial phase can be recognized through the opening/closing of the energy gap -- thus, the in-gap state should exist in a finite system for realizing topological superconductivity. We find a similar pattern in our results. In Fig.~\ref{fig:energy_levels}, we present the energy spectrum of Haldane against $t_{2}/M$ for various cases.

The ``simple'' Haldane model exhibits a topological transition at $t_{2}/M = \pm 1$, which is is visible [Fig.~\ref{fig:energy_levels}(a)]. For large Haldane parameters ($\vert t_{2} / M \vert > 1$), we find that the spectrum becomes gapped. Interestingly, the Haldane model in the presence of superconductivity does not realize a topological superconducting state [Fig.~\ref{fig:energy_levels}(b)]. As shown previously, even if the system is topological, the superconducting states do not necessarily exhibit features similar to a topological superconductor (e.g. gapless edge modes). Finally, the situation changes in Fig.~\ref{fig:energy_levels}(c), where we consider the Haldane model with superconductivity and SOC. In this case, we find gapless states again with a topological phase transition at $ t_{2} / M = \pm 1$, reminiscent of the first case. Note that, the numerical transition points contain small deviations in comparison to theoretical points due to the finite size of the system.

Tuning SOC in the superconducting Haldane model can open/close the energy gap. This can be seen in Fig.~\ref{fig:soc}(a), where we plot the energy spectrum as a function of the SOC strength. To elucidate the momenta contribution, we highlight eigenenergies at K and K’ valleys by red and blue lines, respectively, and keep other states gray. The figure showcases blue in-gaps topological states around zero energy, which originate from the K’ valley. The superconducting Haldane model has no subgap edge states, but we find states between the superconducting gap as soon as the $\lambda_R/t$ is made finite. Such in-gap states eventually touch each other as the coupling
is increased, and the spectrum remains gapless within
a window of $\lambda_R/t$. Beyond this window, the spectrum
becomes gapped again.

Thus, our results highlight the importance of SOC in
obtaining the in-gap topological edge states. This trend
is also visible in Fig.~\ref{fig:energy_levels}(c), where the spectrum contains in-gap states for $t_2>M$. Nevertheless, we can conclude that the condition for the appearance of the in-gap states is a complicated function of the SOC parameter.

\begin{figure}[t]
	\includegraphics[width=\columnwidth]{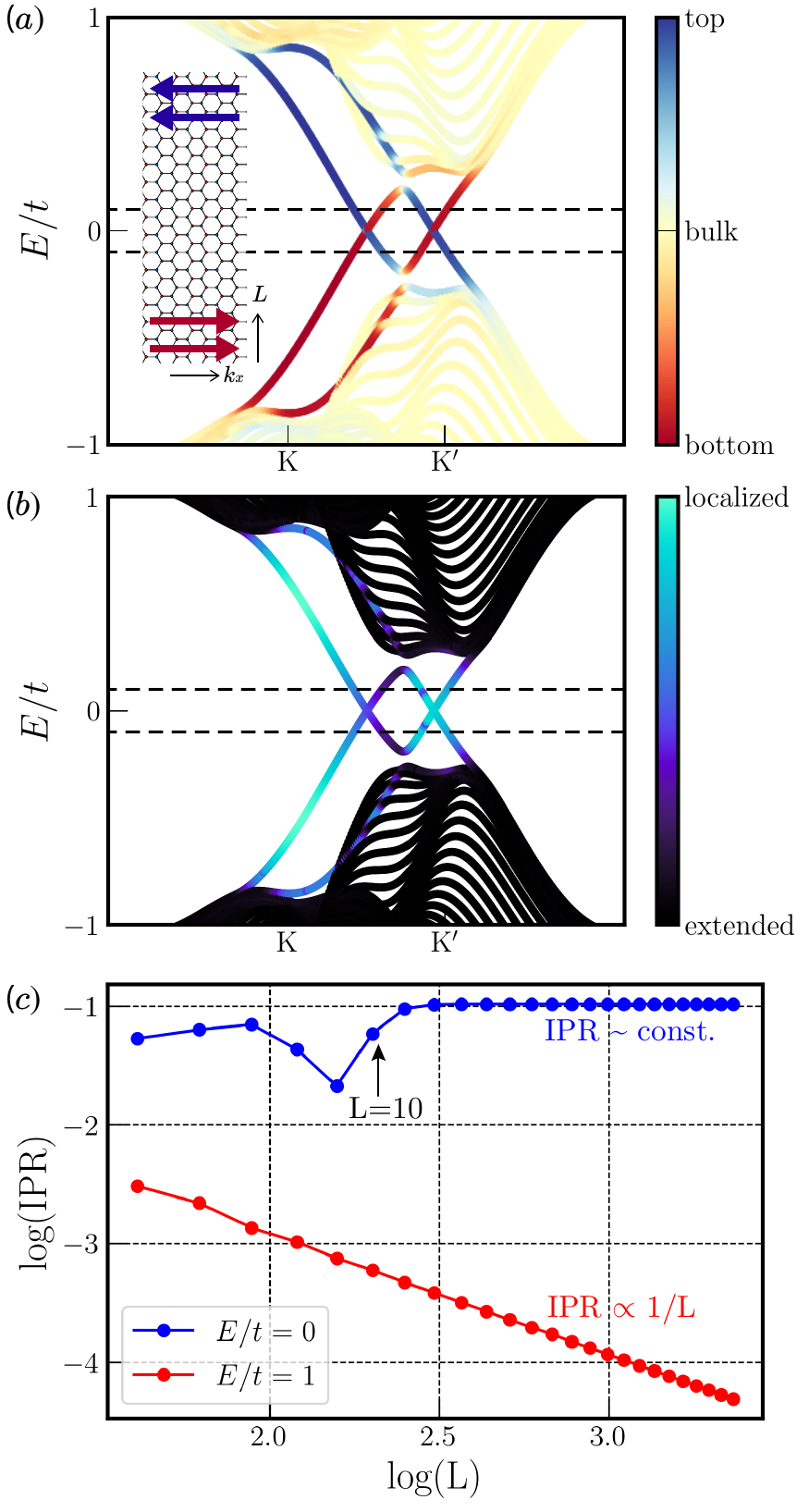}
	\caption{The band structure of the Haldane model is shown in presence of superconductivity and SOC, with the color denoting (a) bulk/edge and (b) localized character of the states. In (a), the inset shows the schematic of a pair of chiral bound states propagating along the boundary of the system. We take a slice at K' and plot the inverse participation ratio in (c) for two energies with varying system sizes in log scale. Results are obtained for $t_2/M=2$, $\lambda_R/t=0.7$, and $\Delta/t=0.1$, $\mu/t=0$.}
	\label{fig:chiral}
\end{figure}

Now we briefly discuss the propagation of the topological in-gap state. To corroborate the chiral nature of the topological states, we project the expectation value of the position operator in real space, and present it in Fig.~\ref{fig:chiral}(a). As expected, most of the states lie outside of the gap exhibiting a bulk character. However, the in-gap states stand out quite well. More precisely, there are two pairs of edge states propagating along the boundary. We find that each pair relates the occupied states at K (K’) with the unoccupied states at K’ (K). Moreover, the pairwise
localizes on the opposite edges. This is a characteristic feature of the chiral superconducting edge modes~\cite{pribiag.beukman.15,kallin.berlinsky.16,zhou.wang.22}, which confirms the realization of the topological superconducting state. 

Here we shall briefly discuss the localization character of the edge states. The localization extent of a wavefunction can be captured	by the measure ``inverse participation ratio''~\cite{anderson.58,kramer.93,wegner.80} (IPR). It is given as the fourth power of the wavefunction
\begin{eqnarray}
\text{IPR}(E) = \sum_i |\psi_i(E)|^4,
\end{eqnarray}
which reflects the number of states that participate in the wavefunction, and gives the inverse of it. Naturally, a small IPR indicates the extended character of the state, while a high IPR means localization. When the IPR is projected on the bands in Fig.~\ref{fig:chiral}(b), we find that the edge states are localized around K and K' valley, with an IPR much higher than the bulk states. Since most wavefunctions participate in an extended state, the IPR typically scales as $1/N$, where $N$ is the number of states. On the contrary, the IPR of a localized state remains constant. With this in mind, we pick an edge ($E/t=0$) and bulk ($E/t=1$) state at K' valley, and compare their IPR in log scale for varying system sizes [see Fig.~\ref{fig:chiral}(c)]. The IPR of the edge state quickly becomes constant, whereas the bulk state IPR scales linearly. This points to the highly localized nature of the edge states. Note that, the constant scaling of the edge state IPR is seen for system sizes $L>10$, below which a fluctuating profile prevails. Therefore, for small system sizes, the chiral edge states at the opposite interfaces can hybridize with each other, potentially destroying the topological phase.  

Now we focus on the role of Rashba SOC in topological superconductivity. We showcase the sublattice-projected band structure evolution for different Haldane parameters $t_{2}/M$ and the aforementioned SOC in Fig.~\ref{fig:soc}. In the topologically trivial regime ($\vert t_2/M \vert < 1$) [Fig.~\ref{fig:soc}(a)], the presence of SOC removes the spin degeneracy of the bands. 
However, the edge states exist outside the superconducting gap. 
The Rashba coupling here cannot move them inside the pairing gap. 
As a result, the spectrum is gapped irrespective of SOC. This trend is different close to the topological phase transition, i.e. $t_2/M \simeq \pm 1$ [Fig.~\ref{fig:soc}(b)] We find that SOC leads to the formation of in-gap states, but in addition, there is a band-touching at K' point. This crossing remains fixed at K' point even if $\lambda_R/t$ is changed. Finally, in Fig.~\ref{fig:soc}(c), we set our attention to the fully topological regime (we set $t_2/M=1.5$). Here we observe the spilling of bands as SOC is turned on. The spectrum features band inversion due to Haldane physics, which highlights the topological nature of edge modes. As $\lambda_R/t$ is increased, the bands split and touch each other at zero energy, eventually forming a pair of zero energy crossings. We also notice that changing the coupling parameter moves the position of the crossings, however, the crossings remain protected for a range of coupling parameters. 

\section{Discussion and experimental realization}
\label{sec:summary}

In this paper, we investigate the topological features of the Haldane model in the presence of onsite superconductivity and Rashba SOC. The Haldane model exhibits a topological phase transition from trivial to nontrivial phase for $t_2 /M = \pm 1$, where $t_2$ is the NNN complex hopping, and $M$ is the staggered flux. The phase transition is associated with two factors: (i) inversion of top and bottom bands and (ii) existence of in-gap states with Chern number $\pm 1$, which cross each other at zero energy. Thus, the system exhibits a gapped spectrum for $\vert t_2/M \vert > 1$.

In contrast, the Haldane model in the presence of superconductivity does not exhibit any in-gap state, irrespective of the Haldane parameter $t_2/M$. However, the band inversion is still preserved for $\vert t_2/M \vert > 1$. This can be treated as an example of a topological system with a ``trivial'' gapped superconducting state.

Finally, we show that the Haldane model with superconductivity and SOC can realize a topological superconducting state. The SOC tuning cannot give rise to in-gap edge modes until
it is in the topological regime of the Haldane model with $\vert t_2/M \vert > 1$. At the critical point and beyond, increasing the SOC makes the edge states touch each other around zero energy, leading to the formation of a gapless superconducting state. Moreover, realized in-gap states exhibit chiral features, where the states propagate along one of the system edges.

Previously, TSC was predicted~\cite{dutreix.guigou.14} on a honeycomb lattice with proximity SC, Rashba SOC, and Zeeman coupling. Here, the magnetic field breaks TRS, and enables the system to enter a Chern phase.
In a realistic scenario,
balancing SC and magnetic field can be challenging since these two are competing phases. One advantage of the Haldane model is that it breaks TRS through complex NNN hopping, which does not hamper the SC state. In fact, orbital current, which are analogous to Haldane flux, are reported~\cite{guguchia.mielke.23} in the superconducting state of the kagome superconductor. Nonetheless, using a setup with Zeeman field produces~\cite{dutreix.guigou.14} two in-gap edge states in the topological phase. In our work, we find four in-gap edge states in comparison. This is a key difference since the bulk topology is related to the number of edge states through the bulk-boundary principle~\cite{hatsugai.1.93,hatsugai.2.93}.

In summary, we show that, in principle, the Haldane model in the presence of superconductivity and SOC can realize a gapless topological superconducting state. Here, topological superconductivity is associated with the realization of the chiral superconducting edge modes. Because of the Haldane model, the realized chiral edge modes can be moved between valleys. Therefore, our results can be a useful guide for experiments involving valleytronics and the realization of superconducting chiral edge modes.

Finally, we dedicate a few words about the possible experimental setups to engineer and measure the proposed edge states. In honeycomb systems, a possible complex next-nearest neighbor hopping can be induced using platforms like ultracold atoms~\cite{jotzu.messer.14}, photonic crystals~\cite{lannebere.silveirinha.19}, adatom deposition~\cite{weeks.hu.11}, and Fe-based oxides~\cite{kim.kee.17}. For example, in Fe-based ferromagnetic insulators, the inversion center forces NN hopping to be real, which is not true for NNN hopping channels. In fact, at NNN sites, the hopping between $d_{yz}$ and $d_{xz}$ orbitals realizes a complex hopping similar to the Haldane model. The setup can be made superconducting using proximity methods like superconducting electrodes~\cite{heersche.herrero.07} or Nb-based heterostructures~\cite{lee.lee.18,moriya.yabuki.20,liu.pawlak.23}, which have been shown to achieve conventional superconductivity in graphene, with a gap in the order of a few meV to few eV. Finally, placing the system on a transitional metal dichalcogenide substrate like $\mathrm{MoS_2}$~\cite{gmitra.fabian.15} can break the mirror symmetry, which leads to Rashba SOC. At the same time, the substrate introduces an effective staggered potential between two sublattices, which is useful for tuning the Haldane model. The scale of the Rashba coupling and staggered potential is of a few meV, and can be tuned through an electric field. With such a backdrop, a possible realization of the proposed edge state in honeycomb systems looks promising.

The topological in-gap edge states are one of the hallmarks of TSC. In-gap states lead to a finite density of states in the superconducting gap of TSC, which is absent in an ordinary {\it s}-wave SC. Scanning tunnelling spectroscopy can track the local density of states by measuring the differential conductance~\cite{kezilebieke.vano.22}. If the in-gap edge state is a Majorana zero mode, the peak conductance at zero bias should be quantized~\cite{lin.sau.12} in terms of $e^2/h$. Additionally, the ratio of in-plane thermal conductance and temperature is predicted to be quantized~\cite{read.green.00} for a Majorana zero mode, which can be verified by thermal interferometry setups~\cite{klocke.moore.22,wei.batra.23,benjamin.das.24}. Scanning tunnelling microscopy is helpful for visualizing the localized edge states, and probe its localization properties.

\begin{acknowledgments}
	We kindly acknowledge support from the National Science Centre (NCN, Poland) under Project No.~2021/43/B/ST3/02166. We gratefully acknowledge the Polish high-performance computing infrastructure PLGrid (HPC Center: ACK Cyfronet AGH) for providing computer facilities and support within computational grant No.~PLG/2023/016835.
\end{acknowledgments}

\bibliography{biblio.bib}

\end{document}